\documentclass[conference]{IEEEtran}
\IEEEoverridecommandlockouts
\usepackage{cite}
\usepackage{amsmath,amssymb,amsfonts}
\usepackage{graphicx}
\usepackage{subcaption}
\usepackage{textcomp}
\usepackage{xcolor}
\usepackage{afterpage}
\def\BibTeX{{\rm B\kern-.05em{\sc i\kern-.025em b}\kern-.08em
    T\kern-.1667em\lower.7ex\hbox{E}\kern-.125emX}}
    
\definecolor{darkBlue}{HTML}{729bba}
\definecolor{green}{HTML}{628c75}
\definecolor{blueGreen}{HTML}{a4decd}
\definecolor{lightGray}{HTML}{d9d9d9}
\definecolor{lightOrange}{HTML}{f0c267}

\usepackage{cleveref} 
\crefname{figure}{Fig.}{Figs.}
\crefname{table}{Tab.}{Tabs.}
\usepackage{tikz}
\usetikzlibrary{positioning,arrows.meta,fit,calc} % used for the efficient working of the positioning system  
\usepackage{eurosym}
\usepackage{mathtools}
\usepackage{MnSymbol}
\usepackage{amsmath}
\usepackage{algorithm}
\usepackage{algorithmicx}
\usepackage{algpseudocode}

\usepackage{float}
\IEEEoverridecommandlockouts\IEEEpubid{\makebox[\columnwidth]{979-8-3503-9678-2/23/\$31.00~\copyright~2023 IEEE \hfill} \hspace{\columnsep}\makebox[\columnwidth]{ }}
\begin{document}

\title{DER Pricing Power in the Presence of Multi-Location Consumers with Load Migration Capabilities}

\author{\IEEEauthorblockN{Sara Mollaeivaneghi, Julia Barbosa, Florian Steinke}
\IEEEauthorblockA{\textit{Energy Information Networks and Systems} \\
\textit{Technical University of Darmstadt}\\
Darmstadt, Germany \\
sara.mollaeivaneghi, julia.barbosa, florian.steinke@eins.tu-darmstadt.de}
}
% and maybe the definition of spatial load shifting and then why is it possible and then literature.
\maketitle

\begin{abstract}
Renewable distributed energy resources (DERs) have the potential to provide multi-location electricity consumers (MLECs) with electricity at prices lower than those offered by the grid using behind-the-meter advantages.
This study examines the pricing power of such DER owners 
in a local environment with few competitors
and how it depends on the MLEC's ability to migrate a portion of the load between locations. 
We simulate a dynamic game between an MLEC and the local DER owners, where the MLEC is modeled as a cost-minimizer and the DER owners as strategic profit maximizers.
We show that, when the MLEC is inflexible, the DER owners' optimal behavior is to offer their electricity close to maximal prices, that is, at the grid price level. However, when the MLEC can migrate a fraction of the load to the other locations, the prices offered by the DER owners quickly decrease to the minimum level, that is, the DERs' grid feed-in tariffs quickly decrease to a lower level, depending on the load migration capability.
%\todo{comment: It also depends on DERs' production capacity. But that is not sth that we examined here.}
\end{abstract}


\section{Introduction}
% (1 Page)
\label{sec:Intro}
With the adoption of renewable energy sources, there is an increasing possibility of renewable distributed energy resources (DERs) to provide sustainable electricity to consumers locally \cite{PEPERMANS2005787}. Local consumers and producers can benefit from behind-the-meter advantages, such as avoiding grid fees. However, since there are typically only few parties behind-the-meter, often only one producer, local price negotiations take place under imperfect market conditions, which may lead to inflated prices for consumers.
Multi-location electricity consumers (MLECs), such as telecommunication or cloud service providers, can potentially benefit from behind-the-meter, local electricity, buying it directly from the DER owners.
However, unlike single-location consumers, the MLECs can use their spatial demand shifting potential to reduce the pricing power of the DER owners. 

Spatial load shifting of the MLECs is commonly called load migration \cite{Fridgen2017}. It has recently become much more practically plausible due to the increasing virtualization of network and computation services  \cite{Vasques2018}.
Load migration between data centers and its effect on demand-side flexibility has been the subject of numerous studies and reviews \cite{Wierman2014, Chen2020}.
It can also be used to provide balancing power  \cite{Fridgen2017,Thimmel2019, Abdelghany2017},
to reduce greenhouse gas emissions \cite{Dandres2017,Zheng2020},
or to save costs on the side of the MLEC by shifting demand to cheaper areas \cite{Qureshi2009, Urgaonkar2011}.
Load migration may at the same time reduce costs for the grid and reduce load variations \cite{Sun2020, Hu2020}.
However, it can also introduce uncertainty in local demand and as a result, utility companies may need to increase electricity prices such that
the MLECs may not benefit from load migration or may only experience minor cost reductions \cite{Camacho2014}. 
While these studies provide valuable insights into the economic benefits of load migration, they all assume that the electricity price is not affected by the MLEC's demand at each location.
However, in reality, the MLEC's demand at each location can have a significant impact on electricity prices and the pricing strategy of local producers.
Competitive behavior between the MLECs and local electric utilities is examined with the help of a two-stage Stackelberg game in \cite{Tran2016}.
However, the work focuses on the effect of flattening the load demand curves and does not analyse local pricing power shift and its dependency on various degrees of load migration capabilities.

This study analyzes the pricing strategies of the local DER owners supplying an MLEC behind-the-meter.
We specifically examine the producers' price setting power in the negotiations with the MLEC and how it depends on the load migration capabilities of the MLEC.

To this end, we simulate a dynamic game between the MLEC and the DER owners.
% how we do it
The MLECs are assumed cost-minimizers, whereas the DER owners strategically maximize their profits using an iteratively estimated price-demand function.
Our findings show that increasing the load migration capability of the MLECs lowers the offered prices by the DER owners, even if the load is not shifted to the other locations.
The findings of this study can help the MLEC operators minimize their overall electricity costs. Additionally it highlights policy and market design requirements for efficient integration of the MLECs and the DERs into the electricity grid.

The remainder of the paper is structured as follows: In Section \ref{sec:Game}, we present a dynamic game that models the interactions between the MLEC and the DER owners, and describe the algorithm developed to determine their decisions. In Section \ref{sec:Case}, we validate our model using exemplary data and investigate the relationship between load migration capability and pricing power of the DER owners. Finally, in Section \ref{sec:Conclusion}, we provide a summary of our findings and offer recommendations for future research.
\section{Dynamic Game: Model and Algorithm}
% (1.5 Pages)
\label{sec:Game}
We use a dynamic game to model the interactions between an MLEC and the DER owners, in which the MLEC and the DER owners are players who interact over multiple sequential periods.

\cref{fig:modelStructure} shows the structure of the proposed model for one iteration of the dynamic game. 
In each iteration of the game, the DER owners first decide about the offered price to sell electricity to the MLEC. Subsequently, the MLEC observes the offered prices from all the DER owners and decides about its electricity purchases from the grid at each location, the electricity purchases amount from each of the DERs, and its total electricity demand at each location.
Finally, the DER owners sell their remaining energy to the grid.
We assume that all players are rational and thus decide on their strategy by solving an optimization problem.
This process repeats over a number of iterations until an equilibrium is reached.

Let $\mathcal{N}=\{1, ..., N\}$ denote the set of locations where the MLEC consumes power.
Moreover, let  $\mathcal{P} = \{1, ..., P\}$ denote the set of the DERs and $\mathcal{P}_n \subseteq \mathcal{P}$ the set of all the DERs based at location $n \in \mathcal{N}$.
Let $n(p)$ denote the location index of the DER $p\in \mathcal{P}$.
The grid's buying price $\pi^{g,b}$ and its location-specific selling prices $\pi^{g,s}_n$ at location $n$ are considered to be fixed over the study period. 
With this notation, in the following, we describe the decision models for the MLEC and the DER owners, as well as the dynamic game.
\afterpage{
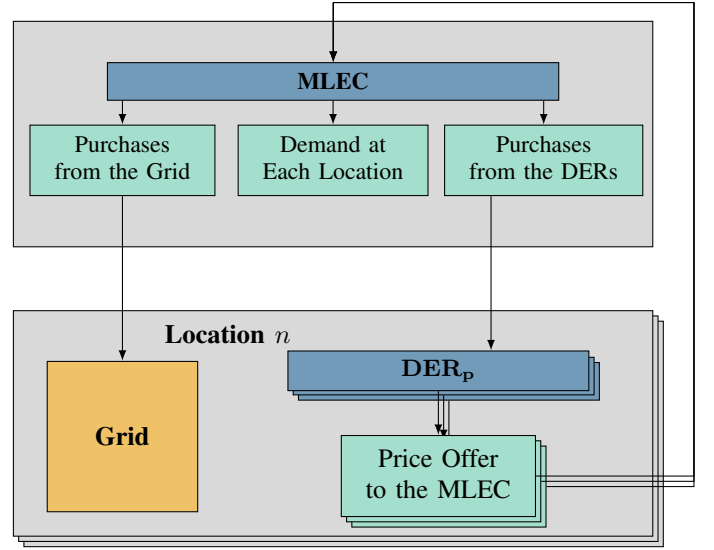
\begin{figure}[t]
  \centering
  % show grid=true
\begin{tikzpicture}[>=latex, scale=0.7]
% First box
\node(box1) at (0,-1) [draw,fill=lightGray,minimum width=8.5cm,minimum height=3cm]{};
\node [font=\small] (mlec) at (0,0) [draw,fill=darkBlue,minimum width=6cm,minimum height=0.5cm] {\textbf{MLEC}};
% \node [font=\small] (prod) at (0,1.5) [fill=lightOrange,minimum width=1cm,minimum height=0.5cm] {\begin{tabular}{c}
%     DERs' \\
%     Offers\end{tabular}};
\node [font=\small] (grid) at (-4,-1.5) [fill=blueGreen,draw,minimum width=1cm,minimum height=0.5cm] {\begin{tabular}{c}
    Purchases \\
    from the Grid\end{tabular}};
\node [font=\small] (demand) at (0,-1.5) [fill=blueGreen,draw,minimum width=1cm,minimum height=0.5cm] {\begin{tabular}{c}
    Demand at \\
    Each Location\end{tabular}};
\node [font=\small] (pr) at (4,-1.5) [fill=blueGreen,draw,minimum width=1cm,minimum height=0.5cm] {\begin{tabular}{c}
    Purchases \\
    from the DERs\end{tabular}};
\draw[->] ([xshift=-4cm]mlec.south) -- ([xshift=-4cm]mlec.south |- grid.north);
\draw[->] (mlec.south) -- (mlec.south |- pr.north);
\draw[->] ([xshift=4cm]mlec.south) -- ([xshift=4cm]mlec.south |- demand.north);
% Second box
\node(box2) at (0.2,-6.7) [draw,fill=lightGray,minimum width=8.5cm,minimum height=3cm]{};
\node(box2) at (0.1,-6.6) [draw,fill=lightGray,minimum width=8.5cm,minimum height=3cm]{};
\node(box2) at (0,-6.5) [draw,fill=lightGray,minimum width=8.5cm,minimum height=3cm]{};
\node (prod-2) [font=\small] at (2.2,-5.7) [draw,fill=darkBlue,minimum width=4cm,minimum height=0.5cm] {};
\node (prod-1) [font=\small] at (2.1,-5.6) [draw,fill=darkBlue,minimum width=4cm,minimum height=0.5cm] {};
\node (prod) [font=\small] at (2,-5.5) [draw,fill=darkBlue,minimum width=4cm,minimum height=0.5cm] {$\mathbf{DER_p}$};
\node (priceOffer-2) at (2.2,-7.7) [fill=blueGreen,draw,minimum width=1cm,minimum height=0.5cm] {\begin{tabular}{c}
    Price Offer \\
    to the MLEC\end{tabular}};
\draw [->] (prod-2.south) -- (priceOffer-2);
\node (priceOffer-1) at (2.1,-7.6) [fill=blueGreen,draw,minimum width=1cm,minimum height=0.5cm] {\begin{tabular}{c}
    Price Offer \\
    to the MLEC\end{tabular}};
\draw [->] (prod-1.south) -- (priceOffer-1);
\node (priceOffer) at (2,-7.5) [fill=blueGreen,draw,minimum width=1cm,minimum height=0.5cm] {\begin{tabular}{c}
    Price Offer \\
    to the MLEC\end{tabular}};
\node (Grid) at (-4,-6.75) [fill=lightOrange,draw,minimum width=2cm,minimum height=2cm] {\textbf{Grid}};
\draw[->] ([xshift=-1cm]pr.south) -- ([xshift=-1cm]pr.south |- prod.north);
\draw [->] (prod.south) -- (priceOffer);
\draw [->] (grid.south) -- (Grid.north);
\coordinate (x1-2) at ([xshift=2.8cm]priceOffer-2.east);
\coordinate (x1-1) at ([xshift=2.9cm]priceOffer-1.east);
\coordinate (x1) at ([xshift=3cm]priceOffer.east);
\coordinate (x2) at ([yshift=9cm]x1.north);
\coordinate (x3) at ([xshift=-6.85cm]x2.west);
\draw (priceOffer-2.east) -- (x1-2) -- (x2) -- (x3) edge[->] ([xshift=0.2]mlec.north);
\draw (priceOffer-1.east) -- (x1-1) -- (x2) -- (x3) edge[->] ([xshift=0.1]mlec.north);
\draw (priceOffer.east) -- (x1) -- (x2) -- (x3) edge[->] (mlec.north);
\node(label) at (-2, -4.8) [black] {\textbf{Location $n$}};
\end{tikzpicture}
\caption{Model Structure}
\label{fig:modelStructure}
\end{figure}
}
\subsection{Decision Model of the MLEC}
The MLEC aims to minimize the total cost of supplying electricity to its different locations. 
The MLEC has a base demand $D_n^{base}$ at each  location $n \in \mathcal{N}$ of which a fraction $\alpha$ can be shifted between locations.

We assume that the base demand $D_n^{base}$ and DERs' production capacities $E_{p}^{gen}$, $p \in \mathcal{P}$, are fixed over the study period. 
We do not take load migration costs into account.
The optimization problem of the MLEC in iteration $i$ of the dynamic game then reads as
\begin{IEEEeqnarray}{lCl}
\label{eq:mlecObj}
\min_{\Theta_{i}} \sum_{n \in \mathcal{N}} \biggl(E^g_{n,i} \pi^{g,s}_n + \sum_{p \in \mathcal{P}_n} E_{p,i} \pi_{p,i}\biggr),
\end{IEEEeqnarray}
subject to
\begin{IEEEeqnarray}{lCl}
    D_{n,i} = E^g_{n,i} + \sum_{p \in \mathcal{P}_n} E_{p,i},\quad & \forall n \in \mathcal{N},\label{eq:demSupBalance-mlec}\\
    \sum_{n \in \mathcal{N}} D_{n,i} = D^{tot}, \label{eq:totDemand}\\
    0 \leq E_{p,i} \leq E_{p}^{gen},\quad & \forall p \in \mathcal{P}, \label{eq:locMarketOffer}\\
    (1-\alpha)D_n^{base} \leq D_{n,i} \leq (1 + \alpha)D_n^{base}, & \forall n \in \mathcal{N},\label{eq:demShiftLimit}\\
    E^g_{n,i} \geq 0,\quad & \forall n \in \mathcal{N} \label{eq:posGridAmount}.
\end{IEEEeqnarray}
The decision variables $\Theta_{i}$ of this problem are the purchased energy from the grid $E^g_{n,i}$ at location $n$, the purchased energy $E_{p,i}$ from each DER $p \in \mathcal{P}$, and the demand at location $n$ $D_{n,i}$.
The MLEC's objective \eqref{eq:mlecObj} is to minimize the total cost of purchasing electricity from the grid and DERs,
where $\pi_{p,i}$ denotes the price of electricity sold by the owner of DER $p \in \mathcal{P}$ to the MLEC.
The constraints ensure the following conditions:
demand-supply balance is maintained at each location \eqref{eq:demSupBalance-mlec}; the total demand $D^{tot}$ of the MLEC is met \eqref{eq:totDemand}; the purchased amount from each DER is less than its capacity $E_p^{gen}$ \eqref{eq:locMarketOffer}; the MLEC's demand at  location $n$ can only deviate by a fraction $\alpha  \in [0, 1]$ from the base demand $D_n^{base}$ at that location \eqref{eq:demShiftLimit};
electricity purchased from the grid $E^g_{n,i}$ at each location is non-negative  \eqref{eq:posGridAmount}.

\subsection{Decision Model of the DER owners}

Each DER owner aims to strategically maximize its profit from electricity sales, i.e., it estimates the reaction of the MLEC to different price offers and chooses its price bids accordingly.

We assume that the DERs are renewable power plants with zero marginal production cost. Therefore, the problem formulation for the DER owner $p$ in iteration $i$ of the dynamic game is 
\begin{IEEEeqnarray}{lCl}
\label{eq:MLECObj}
	\max_{\pi_{p,i}} & \qquad & E_{p,i}(\pi_{p,i})\, \pi_{p,i} + (E_{p}^{gen} - E_{p,i}) \pi^{g,b}, \label{eq:proOpt}
\end{IEEEeqnarray}
subject to
\begin{IEEEeqnarray}{lCl}
E_{p,i}(\pi_{p,i}) = a_{p,i}-b_{p,i}\pi_{p,i},  \label{eq:evalMarktShare}\\
\pi^{g,b}\leq \pi_{p,i} \leq \pi^{g,s}_{n(p)} \label{eq:priceLimits}.
\end{IEEEeqnarray}
The decision variable is the offered energy price to the MLEC $\pi_{p,i}$.
%\todo{Actually, $E_{p,i}$ is also a variable here, which is the expected selling amount to MLEC. Should we mention this?} ==> With this notation is is a functino depending on pi
The objective \eqref{eq:proOpt} is to maximize the profit earned by selling electricity to the MLEC and the rest to the grid. 
The constraints imply that the DER owner estimates its market share depending on the bid price $\pi_{p,i}$ using a linear function \eqref{eq:evalMarktShare} with parameters $a_{p,i}$ and $b_{p,i}$
and that
the plausible DER owner offers are limited by the grid's buying and selling prices \eqref{eq:priceLimits}.

\subsection{The Dynamic Game}

To find the equilibrium between the MLEC and the DER owners, we simulate the negotiation process with \Cref{alg:pseudo}. 

During the iterative negotiation process, each DER owner $p \in \mathcal{P}$ stores its current knowledge about the MLEC's reaction to its possible price offers, i.e., an estimate of the price-demand curve, 
as a set $\mathcal{X}_p$ of pairs of offer prices and the corresponding energy amounts bought by the MLEC. 
The set $\mathcal{X}_p$ is initialized using the following rationale (Line 1):
each DER owner assumes that by offering a price slightly lower (by $\epsilon >0$) than the grid's selling price, it can sell all its generation to the MLEC, and by offering a price slightly higher than the grid's price, it cannot sell anything.
The DER owner also stores the optimal price and the optimal own profit seen so far (Line 2). 

In each iteration $i$, all DER owners first (Line 4-11) estimate the parameters $a_{p,i},b_{p,i}$ of the linear price-demand curve used in \eqref{eq:evalMarktShare} using linear regression.
If the linear regression worked well, i.e., the coefficient of determination $R^2_{p,i}$ of the linear regression is high,
the DER owners decide to choose the next offer price $\pi_{p,i}$ by optimally exploiting their current knowledge $\mathcal{X}_p$.
In case their existing knowledge in inconclusive, i.e., $R^2_{p,i}$ is low,
they try to better explore the MLEC's reaction to different offer prices.
To this end, they choose a new offer price at random from a normal distribution centered around the best price choice so far. The variance of this distribution decreases as the iteration number increases, following a decreasing function $\sigma_i^2 = \sigma_0^2/i$.

Next (Line 12), the MLEC determines the energy amounts to buy from each DER.
%\todo{Why do we use p again in the notation in lines 12 and 19?}
Lastly (Line 13-18), the DER owners update their knowledge about the MLEC's behavior
as well as their storage of the optimal price and profit seen so far.
The actual profit is computed as 
$\text{Profit}(\pi_{p,i},E_{p,i}) = E_{p,i} \pi_{p,i} + (E_{p}^{gen} - E_{p,i}) \pi^{g,b}$ in this step.

The algorithm terminates (Line 19-23) when all DER owners' price decisions change by less than $3 \text{ ct/kWh}$ in consecutive iterations. 
This entire process is repeated ten times, and the results are averaged to mitigate the influence of random factors.

\renewcommand{\tt}[1]{[#1]\,}
\def\NoNumber#1{{\def\alglinenumber##1{}\State #1}\addtocounter{ALG@line}{-1}}

\begin{algorithm}
	\caption{Compute Equilibrium of Price Negotiation}
	\label{alg:pseudo}
	\begin{algorithmic}[1]
		\State 
		$\mathcal{X}_p \gets \{(\pi^{g,s}_{n(p)} - \epsilon, E_p^{gen}), (\pi^{g,s}_{n(p)} + \epsilon, 0)\}$\NoNumber{\hfill\algorithmiccomment{Init Price-Demand Data}}
		\State
		$\pi_p^{opt} \gets \pi^{g,s}_{n(p)}- \epsilon, \; \Pi_p^{Opt} \gets 0, \; i \gets 0$
		\While{True}
		\For{each DER $p$}
		\State $a_{p,i},b_{p,i},R^2_{p,i} \gets \text{LinearRegression}(\mathcal{X}_p)$
		\NoNumber{\hfill\algorithmiccomment{Estimate Price-Demand Curve}}
		\If{$R^2_{p,i} > 0.7$}
		\State 
		$\pi_{p,i} \gets \text{Opt. \eqref{eq:proOpt}-\eqref{eq:priceLimits} using }a_{p,i},b_{p,i}$
		\algorithmiccomment{Exploit}
		\Else
		\State 
		$\pi_{p,i} \gets
		\text{RandomNormal}(\pi_p^{opt},\sigma_i^2)$
		\algorithmiccomment{Explore}
		\EndIf
		\EndFor
		\State 
		$\{E_{p,i}:p \in \mathcal{P}\} \gets \text{Opt. \eqref{eq:mlecObj}-\eqref{eq:posGridAmount} using }\{\pi_{p,i}:p \in \mathcal{P}\}$
		\NoNumber{\hfill\algorithmiccomment{MLEC}}
 		\For{each DER $p$}
 		\State 
 		$\mathcal{X}_p \gets \mathcal{X}_p \cup \{(\pi_{p,i}, E_{p,i})\}$
 		\If{$\Pi_p^{Opt} < \text{Profit}(\pi_{p,i}, E_{p,i})$}
 		\State $\pi_p^{opt} \gets \pi_{p,i}, \;
 		\Pi_p^{Opt} \gets \text{Profit}(\pi_{p,i}, E_{p,i})$
 		\EndIf
 		\EndFor
 		\If{Change of $\{\pi_{p,i}:p \in \mathcal{P}\}$ small}
 		\State terminate
 		\Else
 		\State $i \gets i+1$
 		\EndIf
 		\EndWhile
	\end{algorithmic}
\end{algorithm}

\section{Case Study}
% (1.5 Page)
\label{sec:Case}

In this section, we present a case study to demonstrate our model.
We first examine the offered prices and the MLEC's expenses when an increasing number of DER owners compete in one location. Moving from a monopoly to a competitive market should lower offer prices. We use this assumption as a validation of our model.
In the second experiment we then examine the offer prices when only one DER is present in each location but the MLEC can shift load between the locations to increase competition.

\subsection{Experimental Setup}
To conduct our experiments, we set the MLEC's base demand to $D_n^{base} = 5 \text{ kW}$ and the total demand to $D^{tot} = N\times D_n^{base}$. 
The market prices are chosen as $\pi^{g,b} = 15 \text{ ct/kWh}$ and $\pi^{g,s}_n = 50 \text{ ct/kWh} - 10 \text{ ct/kWh} \times \dfrac{n-1}{N-1}$.
All DERs have $E_p^{gen} = 10 \text{ kW}$, as is typical for decentral PV plants.

The game is iterated until the prices converge, which is defined as a change of no more than $0.05 \text{ ct/kWh}$ between 30 consecutive iterations.
We implement the dynamic game in Python and solve the linear program of MLEC and the quadratic program of each DER using Gurobi \cite{gurobi}.

\subsection{Competition In One Location}
\begin{figure*}[!h]
\centering
\includegraphics[width=0.45\textwidth]{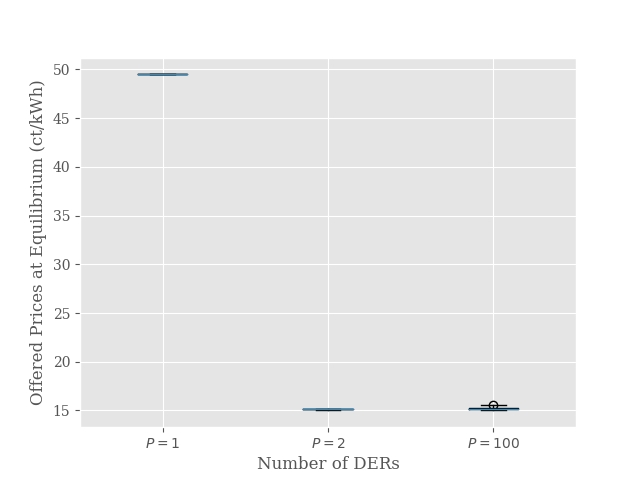}
\quad
\includegraphics[width=0.45\textwidth]{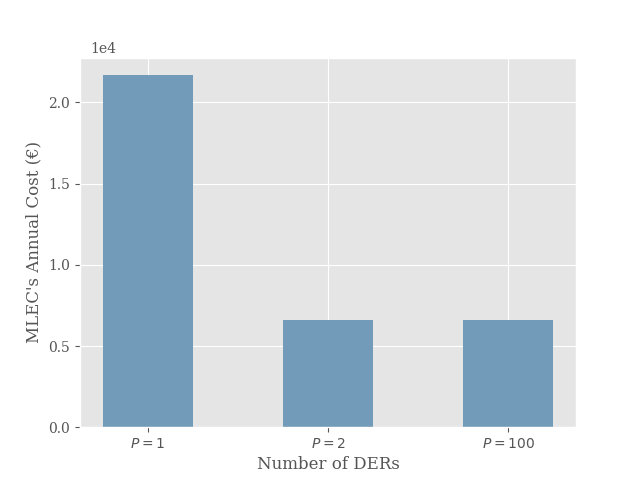}
\caption{Competition between different numbers of DER owners in one location: (left) distribution of final offer prices, (right) MLEC's total costs.
As the capacity of each DER is higher than the demand, even for the case of p=2, the market power breaks and all the producers offer the minimum price at equilibrium.}
\label{fig:SanCheck}
\end{figure*}

The offered prices and the MLEC's expenses for $N = 1$ and $P \in \{1,2,100\}$ are shown in \cref{fig:SanCheck}.
For $P = 1$ the observed offer prices equal the grid's selling price.
Increasing the number of production competitors lowers
the offered prices towards the plausible minimum, i.e., the grid's buying price.
This behavior is as expected since for $P=1$ we have a local production monopoly.
The producers' pricing power decreases when there is more local competition. Here, as the capacity of each DER is higher than the demand, even for the case of p=2, the offered prices at equilibrium are equal to the minimum price.

\bigskip
\subsection{Competition Between Locations}
With these simulations, we investigate the effect of load migration capabilities of the MLEC.
We assume $P = N$, i.e., one DER per location which is realistic for many behind-the-meter microgrids.
We change the MLEC's demand shift capability parameter $\alpha$ in steps of $0.1$.

\begin{figure*}[ht]
\centering
\includegraphics[width=0.4\textwidth]{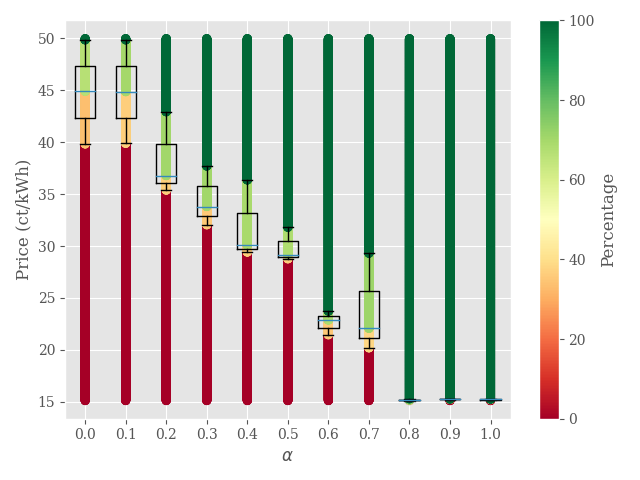}
\quad
\includegraphics[width=0.4\textwidth]{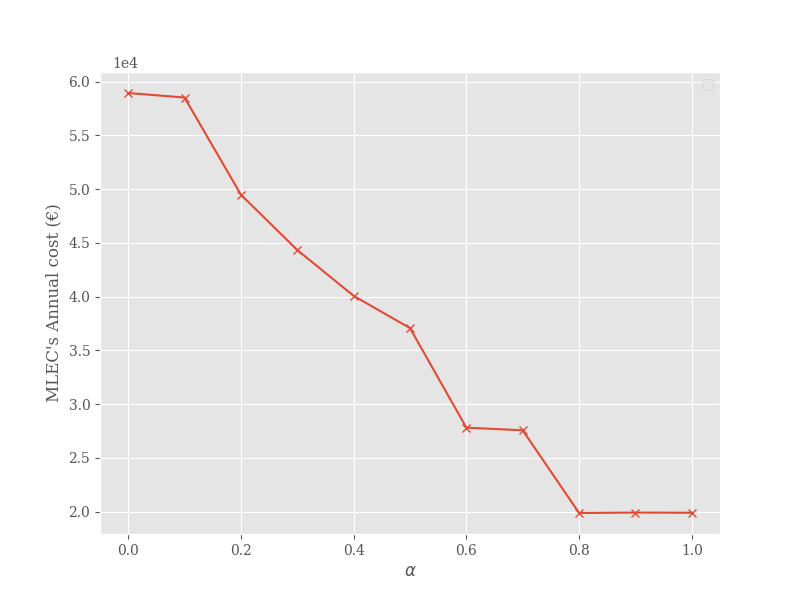}
\caption{Competition between three locations for different levels of MLEC's load migration capability: 
(left) distribution of final offer prices of the DER owners and the percentage of the MLEC's load covered at or below each price level as color-coding.
(right) MLEC's total costs.
Increasing load migration capability reduces offer prices and the total expenses of the MLEC. }
\label{fig:n=3}
\end{figure*}

\begin{figure*}[ht]
\centering
\includegraphics[width=0.4\textwidth]{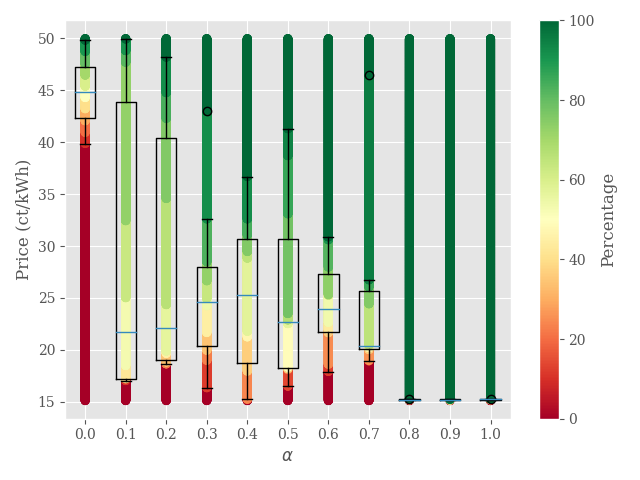}
\quad
\includegraphics[width=0.4\textwidth]{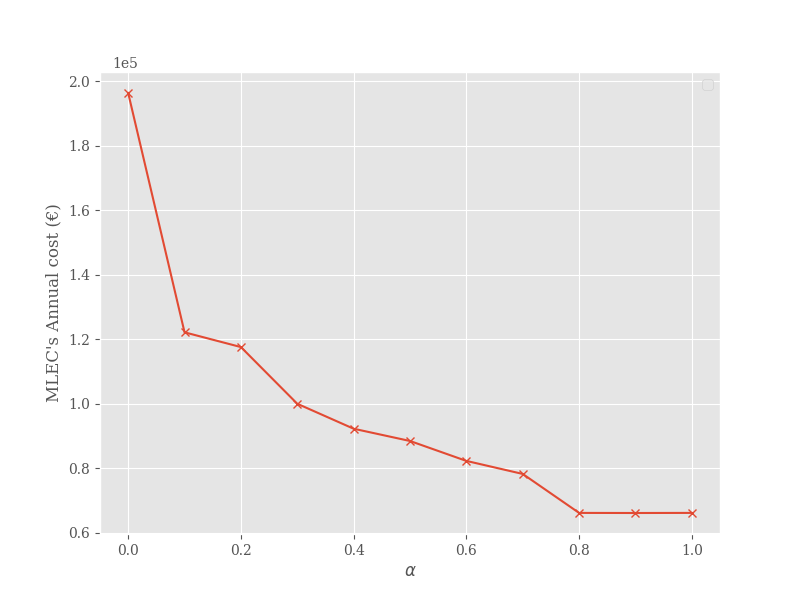}
\caption{Competition between ten locations for different levels of MLEC's load migration capability: 
(left) distribution of final offer prices of the DER owners and the percentage of the MLEC's load covered at or below each price level as color-coding.
(right) MLEC's total costs.
Increasing load migration capability reduces offer prices and the total expenses of the MLEC. 
}
\label{fig:n=10}
\end{figure*}

The resulting offer prices and the MLEC's expenses are shown in \cref{fig:n=3} for $N = 3$ locations and in \cref{fig:n=10} for $N = 10$ locations.

When the value of $\alpha$ is zero, it means that load cannot be shifted between locations. In this case, the prices are at their maximum level, i.e., the grid's selling prices.
In this situation the locations are independent of each other and we observe local production monopolies as in the previous experiment.

However, as the value of $\alpha$ increases and the MLEC gains the ability to shift fractions of the load to cheaper locations, the offer prices decrease. This indicates that load migration can break the pricing power of local producers. The MLEC purchases most of its energy from these cheaper locations, which significantly reduces its electricity bill.  Specifically, we observe a reduction in the MLEC's bill by $65\%$ for $N=3$ and $67\%$ for $N=10$. This implies that MLECs with many locations can particularly benefit from load migration capabilities.

\section{Conclusion}
\label{sec:Conclusion}
In this study, we examined the impact of the MLEC's load migration capability on the pricing power of the local DER owners. The results demonstrate that the load migration capability significantly reduces the pricing power of local producers, resulting in considerable cost savings for the MLEC. Therefore, the MLECs should invest in load migration capabilities, especially those with a larger number of locations.

The study can be extended in several directions. First, the model can be extended to include grid limitations and other consumers in each location. Second, the grid operator can be included as a player in the game to account for the uncertainty in demand prediction caused by the MLEC's load migration, which could have significant effects on the reliability and robustness of the grid.
Third, it's worth noting that our study primarily focuses on the pricing power of local producers. However, given that the MLECs are significant electricity consumers, they have the potential to strategically manipulate local negotiations to their advantage as well. 
Furthermore, the investigation of Deep Reinforcement Learning (DRL) as a potential tool for finding the equilibrium in the interactions between MLEC and DERs can also be suggested as future work.

\bibliographystyle{IEEEtran}
\bibliography{ISGTpaper.bib}
\end{document}